\PassOptionsToClass{natbib=true}{acmart}
\documentclass[sigconf]{acmart}
\AtBeginDocument{%
  }

\copyrightyear{2026}
\acmYear{2026}
\acmConference[SCA/HPCAsia 2026]{SCA/HPCAsia 2026: Supercomputing Asia and International Conference on High Performance Computing in Asia Pacific Region}{January 26--29, 2026}{Osaka, Japan}
\acmBooktitle{SCA/HPCAsia 2026: Supercomputing Asia and International Conference on High Performance Computing in Asia Pacific Region (SCA/HPCAsia 2026), January 26--29, 2026, Osaka, Japan}
\acmDOI{10.1145/3773656.3773658}
\acmISBN{979-8-4007-2067-3/2026/01}

\usepackage{caption, subcaption}
\usepackage{fontawesome5} %
\usepackage{hyperref}
\usepackage{amsmath,amsthm}
\usepackage{algorithmic}
\usepackage{graphicx}
\usepackage{textcomp}
\usepackage{xcolor}
\usepackage{csquotes}

\newtheorem{example}{Example}

\makeatletter %
\renewcommand{\@authornotemark}{}%
\makeatother
\begin{document}

\title{Integrating Quantum Software Tools with(in) MLIR}

\author{
  Patrick Hopf\textsuperscript{$\dagger$ $\ddagger$},
  Erick Ochoa Lopez\textsuperscript{$\S$ *}
  Yannick Stade\textsuperscript{$\dagger$},
  Damian Rovara\textsuperscript{$\dagger$},
  Nils Quetschlich\textsuperscript{$\dagger$},\\
  Ioan Albert Florea\textsuperscript{$\dagger$},
  Josh Izaac\textsuperscript{$\P$},
  Robert Wille\textsuperscript{$\dagger$ $\ddagger$ $\|$},
  and Lukas Burgholzer\textsuperscript{$\dagger$ $\ddagger$}
}
\authornote{Work done while at Xanadu Quantum Technologies Inc., Toronto, Canada.}
\affiliation{
  \institution{\vspace{0.5em} \textsuperscript{$\dagger$}Technical University of Munich, Munich, Bavaria, Germany}
  \institution{\textsuperscript{$\ddagger$}Munich Quantum Software Company (MQSC), Garching near Munich, Bavaria, Germany}
  \institution{\textsuperscript{$\|$}Software Competence Center Hagenberg, Hagenberg, Upper Austria, Austria}
  \institution{\textsuperscript{$\P$}Xanadu Quantum Technologies Inc., Toronto, Ontario, Canada}
  \institution{\textsuperscript{$\S$}AMD, Markham, Ontario, Canada}
  \vspace{0.5em}
  \institution{\{patrick.hopf, yannick.stade, damian.rovara, nils.quetschlich, ioan.florea, robert.wille, lukas.burgholzer\}@tum.de}
  \institution{eochoalo@amd.com, josh@xanadu.ai}
  \city{}
  \country{}
}
\authorsaddresses{}

\renewcommand{\shortauthors}{Hopf et al.}

\begin{abstract}
Compilers transform code into action:
They convert high-level programs into executable hardware instructions---a crucial step in enabling reliable and scalable quantum computation.
However, quantum compilation is still in its infancy, and many existing solutions are ad hoc, often developed independently and from scratch.
The resulting lack of interoperability leads to significant missed potential, as quantum software tools remain isolated and cannot be seamlessly integrated into cohesive toolchains.

The \emph{Multi-Level Intermediate Representation (MLIR)} has addressed analogous challenges in the classical domain.
It was developed within the LLVM project, which has long powered robust software stacks and enabled compilation across diverse software and hardware components, with particular importance in high-performance computing environments.
However, MLIR's steep learning curve poses a significant barrier to entry, particularly in quantum computing, where much of the software stack is still predominantly built by experimentalists out of necessity rather than by experienced software engineers.
\vspace{100pt}
\newpage
This paper provides a practical and hands-on guide for quantum (software) engineers to overcome this steep learning curve.
Through a concrete case study linking {Xanadu}'s PennyLane framework with the Munich Quantum Toolkit (MQT), we outline actionable integration steps, highlight best practices, and share hard-earned insights from real-world development.
This work aims to support quantum tool developers in navigating MLIR's complexities and to foster its adoption as a unifying bridge across a rapidly growing ecosystem of quantum software tools, ultimately guiding the development of more modular, interoperable, and integrated quantum software stacks.
\end{abstract}


\begin{teaserfigure}
  \includegraphics[width=0.95\textwidth]{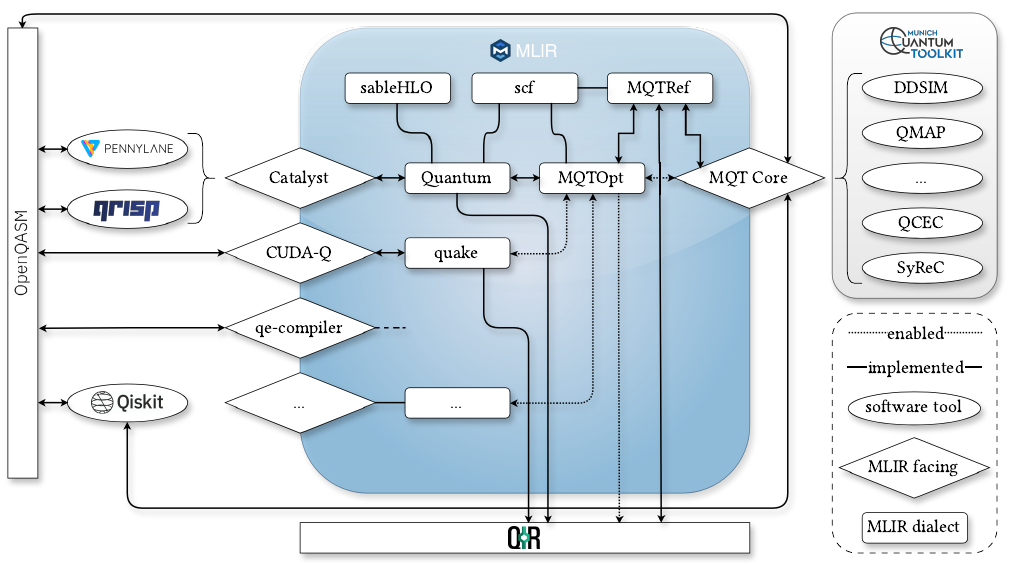}
  \caption{Quantum computing software tools within and outside of the MLIR ecosystem.}
  \Description{Map of quantum computing software tools within and outside of the MLIR ecosystem.}
\end{teaserfigure}

\maketitle

\section{Introduction}

With the rapid growth of quantum software tools and the increasing capabilities of quantum hardware, \emph{quantum compilers} become increasingly important.
Those quantum compilers serve as translators between high-level programming languages and low-level hardware instructions, thereby playing a critical role in connecting two rapidly evolving domains.
On one side, industrial and academic teams are developing specialized quantum software tools optimized for diverse needs; on the other, hardware providers offer a wide range of quantum computing platforms---including superconducting, trapped-ion, neutral-atom, and photonic systems---each with their own native gate set and hardware-specific capabilities and constraints.
To avoid reinventing the wheel each time a new software tool or hardware platform emerges, there is a clear need for an extensible and robust compilation infrastructure that can mediate between the zoo of high-level quantum programming languages and heterogeneous quantum backends---as is taken for granted in classical and HPC systems.

However, compared to the mature state of classical compilation frameworks, tools dedicated to quantum compilation are only emerging slowly.  
Many rely on ad-hoc solutions tailored to narrow use cases, are developed in isolation, and frequently reinvent core components from scratch. 
The resulting lack of interoperability leads to a significant loss of potential, as tools---and their users---could mutually benefit if they were designed to integrate seamlessly.

Fortunately, there is hope:
For classical compilation flows, the \emph{Multi-Level Intermediate Representation (MLIR)}~\cite{mlir} framework has already proven to be an effective solution to many of the challenges currently faced by quantum compilation.  
Adopting this well-established infrastructure for quantum compilers is a natural choice for several reasons:
\begin{itemize}
    \item As part of the LLVM~\cite{Lattner_2004} project, MLIR has been widely adopted---not only by major software frameworks such as JAX~\cite{jax2018github}, TensorFlow~\cite{tensorflow2015}, and PyTorch~\cite{Paszke_2019}, enabling deep integration with classical workflows, but also by hardware vendors such as {Intel} and {NVIDIA}, establishing crucial links to classical and HPC hardware---a key requirement for hybrid quantum-classical programs.
    \item Moreover, MLIR allows for the seamless composition of multiple compilation passes within a shared infrastructure, promoting modularity, reuse, and interoperability.
    \item Ultimately, it is more pragmatic for quantum software developers to extend a robust and ubiquitous classical compiler infrastructure with quantum-specific capabilities, rather than building entirely separate tooling from scratch and incrementally adding classical features such as hardware interfacing, linear algebra, or control flow.
\end{itemize}

However, MLIR is an overwhelmingly complex project, and its steep learning curve significantly hinders adoption---especially in a community that primarily consists of lateral entrants from various disciplines other than computer science or, specifically, compiler design.
Hence, it requires pioneering interdisciplinary projects that connect both worlds: quantum computing on the one side and compiler design on the other side.

\begin{figure*}[t]
    \vspace{-7pt}
    \centering
    \includegraphics{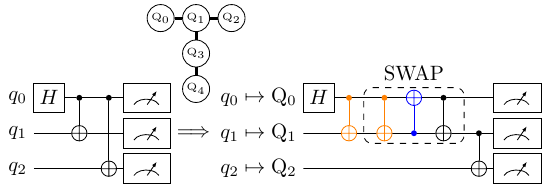}
    \caption{Compilation of a quantum program---preparing the GHZ state---for a \mbox{T-shaped} five-qubit architecture which supports the native gate set $\{\mathrm{H}, \mathrm{T}, \mathrm{CNOT}\}$.}
    \Description{Illustration of a three-qubit GHZ circuit on the left and a T-shaped five-qubit device topology on the right, showing qubit labels, native gates, and inserted swap/cnot operations used during mapping and optimization.}
    \label{fig:init_qc}
\end{figure*}

This paper aims to support developers in navigating the complexities of MLIR and foster its adoption as a unifying bridge across the fragmented quantum software ecosystem.  
To this end, we present a practical guide tailored to quantum software engineers, helping them overcome the steep learning curve of working with MLIR in the context of quantum compilation.  
We demonstrate how MLIR serves as a unified backbone for quantum program representation and transformation workflows, leveraging its lightweight and modular plugin infrastructure.  
A hands-on approach is emphasized throughout, highlighting best practices drawn from a year-long integration effort connecting two major quantum software frameworks: {Xanadu}'s PennyLane~\cite{bergholm_2022} and the Munich Quantum Toolkit (MQT)~\cite{Wille_2024}.

The remainder of this paper is structured as follows:
\autoref{sec:background} reviews relevant concepts in both quantum and classical compilation and provides an overview of related work.  
\autoref{sec:motivation} then outlines the problem of limited interoperability among quantum software tools and motivates MLIR as a potential solution, whose adoption is currently limited by the steep learning curve associated with the framework.
To overcome the high entry barrier, \autoref{sec:casestudy} presents best practices and a detailed case study demonstrating the integration of two representative quantum software frameworks---PennyLane and MQT---within the MLIR ecosystem.  
In \autoref{sec:discussion}, this is followed by a discussion of key insights and future prospects arising from such an integration.
Finally, \autoref{sec:conclusion} summarizes the contributions of this work.

\section{Background}\label{sec:background}

To keep this paper self-contained, this section briefly reviews the fundamentals of quantum and classical compilation and concludes with an overview of related work on intermediate representations.

\subsection{Quantum Compilation}

To execute a quantum program on a chosen quantum device, the program must first be converted into an executable.
This process is conducted by \emph{compilers} that transform the quantum program to a sequence of hardware-specific instructions (in rough analogy to a classical instruction set architecture, or ISA).
Those instructions must adhere to the constraints induced by the chosen quantum computer.

To this end, the \emph{compilation} is typically divided into compilation \emph{passes} that operate on a quantum circuit, which consists of qubit wires and gate operations.
These passes can be executed sequentially or invoked multiple times, resulting in complex compilation \emph{flows}.
All passes can be broadly classified into the following three categories:
\begin{itemize}
  \item \emph{Placement and Routing}: 
  These passes assign each qubit present in the quantum program to a physical qubit on the target hardware.
  They ensure that all gates are executable with respect to the device's topology.  
  To resolve topological constraints, additional operations---such as SWAP gates on superconducting systems or move/shuttling operations on neutral atom and trapped-ion platforms---may be introduced, as demonstrated in~\cite{Wille_2023, schmid2024hybrid, Kreppel_2023, Schoenberger_2024, bach2025}. %

  \item \emph{Synthesis}: These passes decompose quantum operations that are not natively supported by the target device into sequences of gates from its native gate set, using strategies such as those proposed in~\cite{Arrazola_2019, 1997Kitaev, younis2021qfast, pehamDepthoptimalSynthesisClifford2023, gilesExactSynthesisMultiqubit2013}. %
    
  \item \emph{Optimization}: These passes aim to improve the efficiency of a quantum program and thereby mitigate some of the inherent limitations of current noisy quantum computers, employing techniques such as those proposed in~\cite{niu2023powerfulquantumcircuitresizing, Hopf_2025, Itoko2019, H_ner_2020, Niu_2024}. %
\end{itemize}

\begin{example}\label{ex:init_compilation}
Assume the quantum program preparing a GHZ state with three qubits shown on the left-hand side of \autoref{fig:init_qc} shall be mapped onto a T-shaped five-qubit architecture (as depicted at the top) and synthesized to its native gate set $\{\mathrm{H}, \mathrm{T}, \mathrm{CNOT}\}$.
By choosing a one-to-one mapping between the circuit and the device qubits and inserting a single SWAP gate---synthesized as three CNOT gates---a fully executable circuit that matches the device's topology can be derived.
Subsequent optimization passes may identify that the two orange \(\mathrm{CNOT}\) gates cancel, and, assuming initial qubit states of \(|0\rangle\), the blue \(\mathrm{CNOT}\) can likewise be removed---leaving only the two trailing \(\mathrm{CNOT}\) operations.
\end{example}

To construct such compilation flows, all constituent compilation passes usually consume and produce the quantum program in a common format---the so-called \emph{intermediate representation}~(IR).
To this end, many quantum software tools have introduced their own IRs at both high and low levels of abstraction; examples include Qiskit's \texttt{DAGCircuit}~\cite{Javadiabhari_2024}, TKET's \texttt{Circuit}~\cite{tket}, TKET2's \texttt{HUGR}~\cite{kochhugr}, as well as MQT's \texttt{QuantumComputation}~\cite{Burgholzer_2025} and PennyLane's \texttt{QuantumTape}~\cite{bergholm_2022}.
While this approach works well \emph{within} an individual tool, it complicates the integration of different tools in a unified compilation flow as demonstrated in \autoref{ex:predictor}.
Furthermore, no shared framework or standardized infrastructure is currently established to facilitate efficient IR translation across different quantum software tools for the purpose of compilation. 

\begin{example}\label{ex:predictor}
The MQT Predictor~\cite{quetschlich_2025} enables the composition of compiler passes from both Qiskit and TKET to construct optimized compilation flows that surpass the capabilities of the individual tools.  
While the reinforcement learning-based interleaving of various passes yields highly optimized quantum programs, the underlying compilation routine---and consequently the model training process---must continually translate between the distinct circuit representations employed by Qiskit and TKET, posing a significant runtime overhead.
\end{example}

\subsection{Classical Compilation}

In contrast to the still-maturing quantum software ecosystem, classical compilation has benefited from decades of development.  
A key success factor has been the use of shared frameworks, most notably LLVM~\cite{Lattner_2004}, a widely adopted compiler infrastructure offering reusable libraries and tools.  
LLVM enables the translation of high-level languages (such as C/C++, Rust, or Java) into low-level, hardware-specific instructions (e.g., assembly) using a unified IR known as LLVM IR.  

Beyond language translation, LLVM has become a cornerstone of modern HPC software stacks.  
It provides the foundation for widely used production compilers such as Clang, supports vectorization and parallelization techniques critical to HPC workloads, and serves as a backend for domain-specific languages and accelerator toolchains.  
By offering reusable optimization passes, target-specific backends, and a well-defined IR, LLVM enables a high degree of portability and performance tuning across diverse HPC architectures, ranging from multi-core CPUs to GPUs and other specialized accelerators, among which quantum processors are beginning to emerge~\cite{Elsharkawy25}.

While LLVM IR facilitates modular compiler design, relying on a single, static IR can make it difficult to express high-level abstractions or low-level hardware details.
This gap led to the development of the \emph{Multi-Level Intermediate Representation (MLIR)}~\cite{Lattner_2020}, a more flexible infrastructure that supports distinct levels of abstraction and is designed for domain-specific compiler engineering.

MLIR allows developers to define custom IRs via dialects and compose tailored passes for their specific domains.
It has been widely adopted by classical software tools such as JAX~\cite{jax2018github}, TensorFlow~\cite{tensorflow2015}, and PyTorch~\cite{Paszke_2019}.
The framework provides native support for packaging and distributing custom dialects and compiler passes as plugins.  
By chaining together multiple passes, developers can construct flexible and modular compilation pipelines.

Internally, MLIR follows Static Single Assignment (SSA) semantics: Meaning each \emph{value} is defined exactly once.
All values are explicitly linked to the \emph{operation} that defines them, as well as to all operations that use them.  
This so-called \emph{definition-use} chain enables efficient program traversal in compilation passes.
MLIR's graph-like program representation---comprising operation nodes and value edges---aligns naturally with quantum circuits, which are often modeled as directed acyclic graphs.

Taken together, these features make MLIR a compelling foundation for representing and transforming quantum programs via custom dialects and pass logic, ultimately enabling the deployment of stand-alone plugins.

\begin{figure*}[t]
  \centering    
  \includegraphics[width=\linewidth]{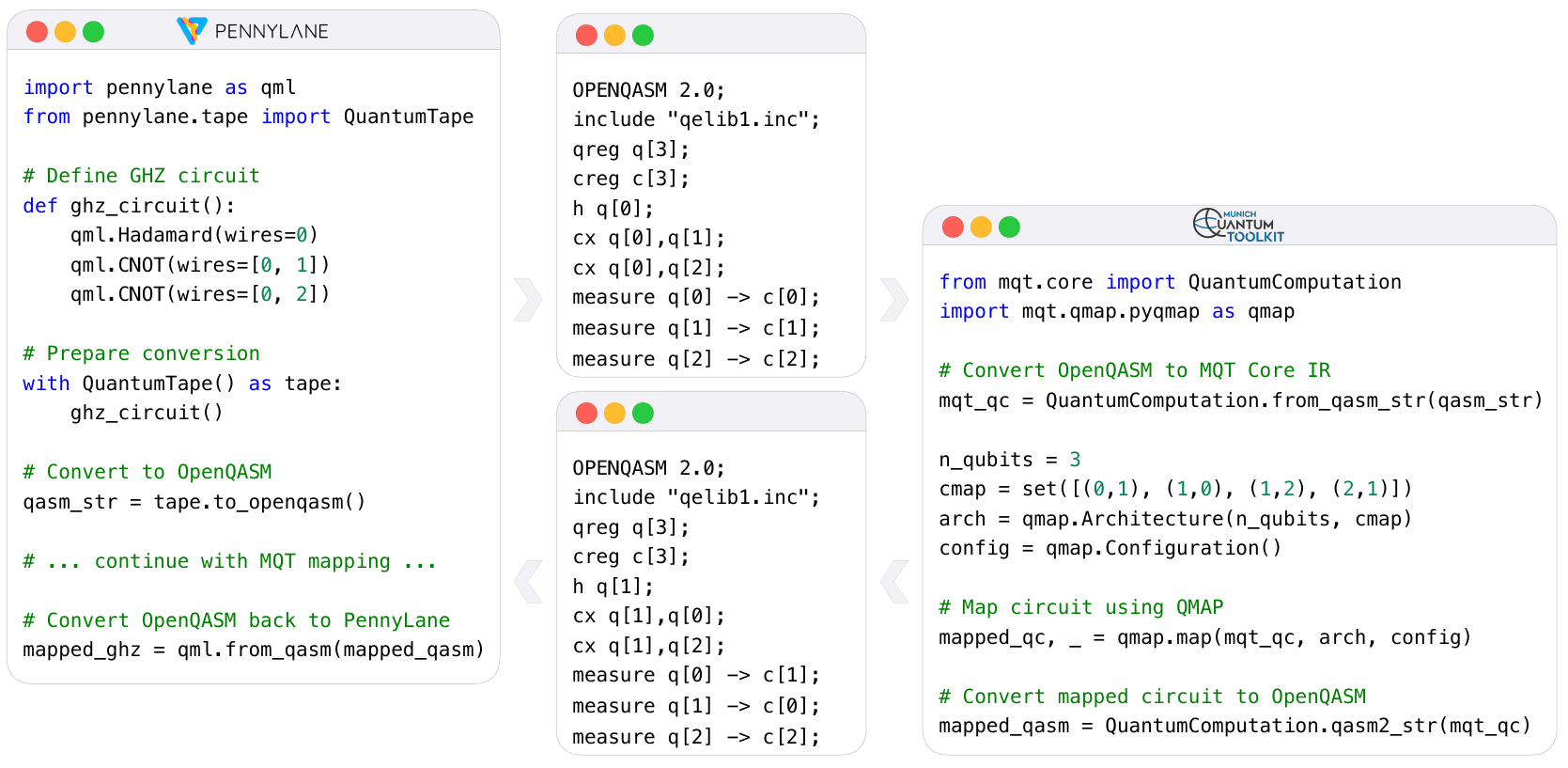} 
  \caption{Mapping a quantum program defined in PennyLane~\cite{bergholm_2022} with MQT QMAP~\cite{Wille_2023} without MLIR integration.}
  \Description{Workflow diagram showing the PennyLane to MQT QMAP mapping that uses OpenQASM and intermediate conversions, highlighting multiple tool boundaries and extraneous translation steps.}
  \label{fig:current_workflow}
\end{figure*}

\subsection{Related Work}

Given the plethora of quantum software tools available today, it is unsurprising that an equally diverse set of quantum program representations has emerged, each tailored to specific needs and use cases.
A variety of data structures have been proposed to describe the functionality of quantum circuits, most notably tensor networks~\cite{Pirvu_2010} and decision diagrams~\cite{Wille_2022, Wille2023}, which aim to represent the exponentially large state space acted upon by quantum operations and are therefore useful for classical quantum circuit simulation~\cite{Hauschild_2018, Zulehner_2019, Grurl2025} or verification~\cite{sander2024equivalence}.  
Other representations are tailored to specific compilation tasks, such as the ZX-calculus~\cite{Kissinger_2020} and its derivatives~\cite{Matthew_2014, vandeGriend_2025}, the stabilizer formalism for Clifford circuits~\cite{Gottesman_1997}, and graph-based models~\cite{wang2022quest, venturelli2019quantum, bandic2023interaction}.  
However, these representations are not designed to serve as general-purpose exchange formats for quantum programs, as they typically prioritize compactness or tractability of specific features over broad expressiveness and interoperability.

Early efforts toward a universal and extensible IR that unifies quantum and classical computation have been introduced with {IBM}'s OpenQASM~3~\cite{Cross_2022}.
It has become widely adopted within the quantum computing community, building on the success of its predecessor, OpenQASM~2~\cite{cross2017open}.  
However, it necessitates reimplementing many concepts that have long been established and refined in classical compiler infrastructures~\cite{stade2024supportingqirthoughtsadopting}.  

To address this limitation, the Quantum Intermediate Representation (QIR)~\cite{QIRSpec2021}---originally proposed by {Microsoft}---has begun to emerge slowly~\cite{stade2024supportingqirthoughtsadopting}.
QIR is built on top of the classically established LLVM framework and extends it with a universal interface between quantum programming languages and various quantum hardware backends. 
To this end, it defines a set of protocols for expressing quantum programs in a hardware- and language-agnostic format embedded within LLVM IR and has been adopted in compilation tools such as~\cite{Wong_2024, Gupta_2025}.
As it is based on LLVM, QIR faces the same challenges encountered in classical compilation.  

Consequently, it is not surprising that MLIR---offering customizable abstraction levels and an extensible IR framework---has recently gained popularity within the quantum computing community.
Q-MLIR~\cite{McCaskey_2021} explores the integration of quantum constructs into the MLIR ecosystem in a manner that remains compatible with QIR interface standards.
Similarly, QIRO~\cite{Ittah_2022} extends MLIR to support quantum-classical co-optimization.
Industry frameworks such as {NVIDIA}'s CUDA-Q~\cite{CUDAQ} and {Xanadu}'s Catalyst~\cite{Ittah_2024} also build on MLIR, providing extensible compiler backends and plugin access for third-party tool developers.

A comprehensive overview and comparative study of initial quantum IRs (such as, e.g., QSSA~\cite{Peduri_2022}, XACC\cite{McCaskey_2020}, or QBIR~\cite{Luo_2020}) as well as more recent MLIR-based approaches (such as Q-MLIR~\cite{McCaskey_2021} and QIRO~\cite{Ittah_2022}) can be found in~\cite{Cardama_2025}.

Ultimately, despite initial efforts toward MLIR-based IRs for quantum compilation, a unified integration path for the diverse set of quantum computing tools remains absent. 
The compiler and quantum communities remain largely disconnected---mainly due to the substantial expertise required in both domains.

\textbf{To unlock the full potential at the intersection of these disciplines, it is essential to make MLIR more accessible and usable for the quantum computing community.}

\section{Motivation}\label{sec:motivation}

Although MLIR provides a powerful infrastructure for building sophisticated quantum compilers capable of connecting the diverse ecosystem of quantum computing tools, its adoption within the community remains limited.
This section examines the key obstacles hindering the broader adoption of MLIR within the quantum community and outlines how this work aims to address them.

\subsection{Considered Problem}

In classical and HPC computing, compilation has become increasingly standardized and unified, with LLVM IR playing a central role.  
The introduction of MLIR and its support for custom dialects has relaxed the constraints of using a single IR in a controlled and coherent way.  
In contrast, the variety of IRs used in quantum computing lacks a unified and shared infrastructure suitable for compilation.
A major reason for that is that quantum computing---and more specifically, quantum software engineering---is still a young area of research without established standards.
The quantum devices themselves have often been the primary focus area of many researchers and hardware vendors, while the software was treated as an afterthought.
This has resulted in a highly heterogeneous quantum computing software landscape that is characterized by a wide variety of quantum software tools. 
These not only differ in their underlying data structures, compilation strategies, and development philosophies, but they also suffer from limited interoperability due to the absence of a shared IR---effectively prohibiting the combination of different tools without significant effort.
This is exacerbated by the rapid pace of the still-young quantum (software) ecosystem and the maintenance and continuous development demands that result from the high volatility.

While there exist \enquote{workarounds} made possible by early standardization efforts (such as OpenQASM), these formats are better suited as human-readable, static program descriptions rather than as dynamic, in-memory representations suitable for compiler infrastructure.
Consequently, they are not typically used as the internal representation in modern quantum software stacks, which leads to significant parsing overhead and the frequent loss of structural and semantic information during translation.

\begin{figure*}[t]
  \vspace{-2pt}
  \centering
  \includegraphics[width=0.5\linewidth]{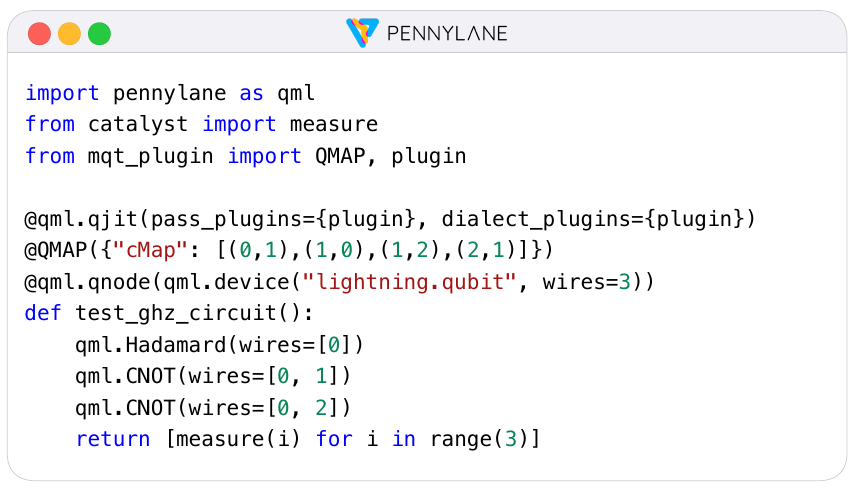}
  \caption{Mapping a quantum program defined in PennyLane~\cite{bergholm_2022} with MQT QMAP~\cite{Wille_2023} using the MLIR plugin system.}
  \Description{Diagram of an MLIR-based plugin workflow that directly converts PennyLane programs into MQT tools via MLIR dialects, eliminating OpenQASM roundtrips and intermediary tool dependencies.}
  \label{fig:plugin-mapping}
\end{figure*}

\begin{example}\label{ex:current_workflow}
    Assume that the quantum program described in \autoref{ex:init_compilation}, which prepares a GHZ state and is shown on the left-hand side of \autoref{fig:init_qc}, is defined using {Xanadu}'s PennyLane~\cite{bergholm_2022} and should be mapped with MQT's QMAP~\cite{Wille_2023}.
    Since the tools have no shared IR, there is no trivial solution to implement the joint compilation flow.
    Therefore, a workaround must be taken as visualized in \autoref{fig:current_workflow}.

    After defining the program as a PennyLane \texttt{QuantumTape}, it must first be exported as an OpenQASM string.  
    This string can then be converted into MQT Core's IR, \texttt{QuantumComputation}, and subsequently mapped using QMAP.
    Once the mapping is complete, the compiled circuit must be converted back through all intermediate steps to make it usable in PennyLane again. 
    Ultimately, this process requires an installation of Qiskit, which is used by \texttt{pennylane-qiskit} for the import of {OpenQASM}---introducing yet another dependency (with its own IR and conversion routines). 

    Furthermore, the qubit assignment resulting from the mapping process is not explicitly preserved in the OpenQASM IR. 
    Thus, there is no way to infer which device qubit a particular qubit from the original circuit has been mapped to, which might be crucial for reasoning about the correctness of the compiled circuit~\cite{burgholzerVerifyingResultsIBM2020}.
\end{example}

Instead of reinventing the wheel by developing yet another IR for each new quantum software tool and relying on cumbersome workflows to communicate between them, decades of experience in classical compiler and software infrastructure can be leveraged.
The MLIR framework has demonstrated significant benefits in the classical domain and offers a robust foundation for extensible and reusable quantum compiler design.
However, understanding and effectively using such a complex and comprehensive framework---developed in C++---is far from trivial and comes with two main difficulties.

The first is technical: Quantum software engineers, often coming from a physics or engineering background, are typically more familiar with Python, as reflected in the predominantly Python-based quantum computing software landscape.  
This results in a natural barrier to entry, as learning MLIR often requires becoming proficient in C++ as well. 
MLIR's heavy use of template programming patterns only further complicates the situation.
Even just setting up the large LLVM project---of which MLIR is a subcomponent---can be a challenge in its own right.  
These projects involve millions of lines of code, come with many dependencies, and rely on \href{https://cmake.org}{CMake} for configuration and building---another powerful but intricate tool that even computer scientists can find challenging to master.

The second is conceptual: even when technical barriers are lowered, MLIR introduces compiler concepts---such as SSA semantics, IR transformations, and pass pipelines---that may be unfamiliar to physicists more accustomed to unitary matrix representations, pattern matching, and physical design aspects.  
While projects like \texttt{xDSL}~\cite{Fehr_2025} aim to reduce this friction by exposing MLIR-like abstractions through Python, the learning curve remains steep for those without a background in compiler construction.

We believe that these factors significantly contribute to the fact that MLIR has not yet seen widespread adoption within the quantum computing community.
Its largely untapped potential to interconnect the fragmented landscape of quantum software tools underscores the need for a clear and accessible guide to help quantum engineers integrate this well-established classical framework into their tools.

\subsection{Contribution}

This paper aims to provide such a practical guide for quantum software developers aiming to harness the potential of MLIR as a compilation framework.
It demonstrates how MLIR can be effectively leveraged to bridge heterogeneous quantum software tools through a shared compilation backbone.

To this end, we present a case study showcasing the integration of two major quantum software frameworks: {Xanadu}'s PennyLane and the Munich Quantum Toolkit (MQT).
By providing the foundational components required for a successful integration within the MLIR ecosystem, this demonstration focuses on only the most essential concepts while also pointing to valuable resources for more in-depth exploration---helping readers get started without being overwhelmed by the full intricacy of MLIR.
We demonstrate a lightweight and modular approach by leveraging MLIR's dedicated plugin infrastructure, thereby minimizing complexity and external dependencies---common barriers to widespread adoption.
A hands-on approach is emphasized throughout, with best practices highlighted and key insights shared from our integration efforts.
To further lower the entry barrier and facilitate broader adoption across the community, the full code is made publicly available as part of the \href{https://github.com/munich-quantum-toolkit/}{MQT repository}~\cite{Wille_2024}.

The following section demonstrates how to achieve the seamless interoperability enabled by MLIR's extensible plugin architecture.

\begin{example}\label{ex:proposed_workflow}
    The workflow described in \autoref{ex:current_workflow} is obviously not optimal. 
    However, using the MLIR plugin system, the complexity of such a workflow is significantly reduced as shown in \autoref{fig:plugin-mapping}.
    Behind the scenes, the program is efficiently converted within MLIR using dedicated dialects and passes.
    Providing the custom plugin and pipeline in  PennyLane is enough to trigger the compilation---avoiding any OpenQASM parsing, dumping, and loading efforts, as well as fully abandoning the Qiskit dependency.
\end{example}

\section{Integration}\label{sec:casestudy}

This section presents a practical integration effort that enables interoperability between two major quantum software frameworks using MLIR. 
The first two parts introduce the individual frameworks, while the third outlines the key components necessary to facilitate their integration within the MLIR ecosystem.
Together, these present a practical blueprint for quantum software engineers seeking to adopt MLIR in their own tooling.

\subsection{{Xanadu}'s PennyLane}

PennyLane~\cite{bergholm_2022} is a cross-platform Python library for quantum programming, with a key differentiator compared to other quantum software frameworks being built-in support for end-to-end autodifferentiation through quantum and classical instructions. Recently, PennyLane has added support for quantum just-in-time (QJIT) compilation via the Catalyst compiler~\cite{Ittah_2024}. Inspired by similar approaches in the Python ecosystem (including frameworks such as Numba and JAX), quantum just-in-time compilation allows dynamic, hybrid quantum-classical programs to be written in Python, but not executed by the Python interpreter. Instead, when executed from Python, the program is captured (including all classical processing, quantum instructions, and control flow), and compiled to an optimized machine binary via MLIR and LLVM. On subsequent calls to the Python program, this pre-compiled binary is instead executed on the specified simulator or hardware device, without the need to recompile or re-capture the program.

Internally, the Catalyst compiler consists of three main components:

\begin{enumerate}
    \item \textbf{A Python frontend}. When the user executes a function to be QJIT-compiled, it is executed with \textit{tracers} representing abstract function parameters, allowing the function's behaviour to be captured. This frontend extends JAX's~\cite{jax2018github} tracing infrastructure to capture Python functions that contain classical instructions, PennyLane quantum operations, and native Python control flow. The frontend then lowers the program representation to MLIR, using a bespoke quantum dialect for the quantum operations, and the StableHLO dialect by OpenXLA for the classical instructions. 
    
    \noindent The bespoke MLIR Quantum Dialect provided by Catalyst allows users to denote their own custom gates, which would be lowered to a runtime function call passing the name of the gate as a parameter. The quantum operations are expressed in SSA form, where a quantum operation takes input qubits and returns output qubits.
    \item \textbf{An MLIR compiler}. During compilation, Catalyst optimizes the program representation to LLVM and compiles the program to a binary. During this process, quantum optimizations are applied directly to the structured program (that is, without removing or unrolling any classical control flow).
    \item \textbf{A runtime}. During execution of the compiled binary, Catalyst will execute quantum instructions on user-specified quantum hardware or simulator devices, such as PennyLane's Lightning simulator suite~\cite{asadi2024}.
\end{enumerate}

\subsection{The Munich Quantum Toolkit (MQT)}

The \emph{\href{https://mqt.readthedocs.io}{Munich Quantum Toolkit (MQT)}}~\cite{Wille_2024} is a collection of \mbox{open-source} software tools for quantum computing developed by the \href{https://www.cda.cit.tum.de/}{Chair for Design Automation} at the \href{https://www.tum.de/}{Technical University of Munich} as well as the \href{https://munichquantum.software}{Munich Quantum Software Company (MQSC)}. 
Among others, it is part of the \href{https://www.munich-quantum-valley.de/research/research-areas/mqss}{Munich Quantum Software Stack (MQSS)}~\cite{sca_hpcasia_2026_1} ecosystem, which is being developed as part of the \href{https://www.munich-quantum-valley.de}{Munich Quantum Valley (MQV)} initiative.
Its overarching objective is to provide solutions for design tasks across the entire quantum software stack. This entails high-level support for end users in realizing their applications, efficient methods for the classical simulation, compilation, and verification of quantum circuits, tools for quantum error correction, support for physical design, and more. These methods are supported by corresponding data structures (such as decision diagrams or the ZX-calculus) and core methods (such as SAT encodings/solvers).

Given the need throughout the entire toolkit to effectively represent and manipulate quantum circuits and the fact that the origins of the project date back to a time when OpenQASM 2 had just been proposed and none of the previously discussed tooling had been around, the MQT features its own C++-based \texttt{QuantumComputation} IR, which is part of the MQT Core~\cite{Burgholzer_2025} library and referred to as \emph{MQT Core IR} in the following.

The IR is used in various top-level libraries throughout the MQT. 
One of these is MQT QMAP~\cite{Wille_2023}.
Its name originates from the fact that it was mainly a tool for mapping quantum circuits to superconducting architectures with a limited connectivity (i.e., solving the placement and route problem). 
Over the years, it has grown to a collection of compilation tools that allows mapping quantum circuits to various qubit technologies, now also including neutral atom quantum computers~\cite{schmid2024hybrid}.

As of this work, MQT Core additionally provides dedicated MLIR dialects that are available starting with version 3.3.3.
\begin{itemize}  %
  \item[\faGithub] \small{\href{https://github.com/munich-quantum-toolkit/core/tree/v3.3.3}{https://github.com/munich-quantum-toolkit/core/tree/v3.3.3}}
  \item[\faBook] \small{\href{https://mqt.readthedocs.io/projects/core/en/v3.3.3/}{https://mqt.readthedocs.io/projects/core/en/v3.3.3}}
\end{itemize}
Such dialects are crucial for the plugin, with one particular dialect, namely \texttt{MQTOpt}, discussed in the following.

\begin{figure*}[t]
  \centering
  \begin{subfigure}{0.48\linewidth}
      \centering
      \includegraphics[width=0.93\linewidth]{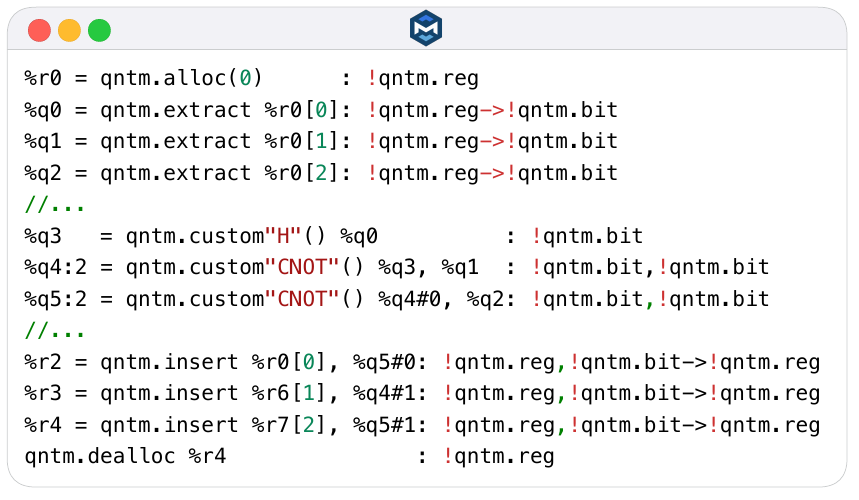}
      \caption{Catalyst \texttt{Quantum} MLIR dialect}
      \Description{Subfigure showing an SSA-style representation of a GHZ circuit expressed in the Catalyst Quantum MLIR dialect, with operations and value names.}
      \label{fig:catalyst-quantum}
  \end{subfigure}
  \hfill
  \begin{subfigure}{0.48\linewidth}
      \centering
      \includegraphics[width=\linewidth]{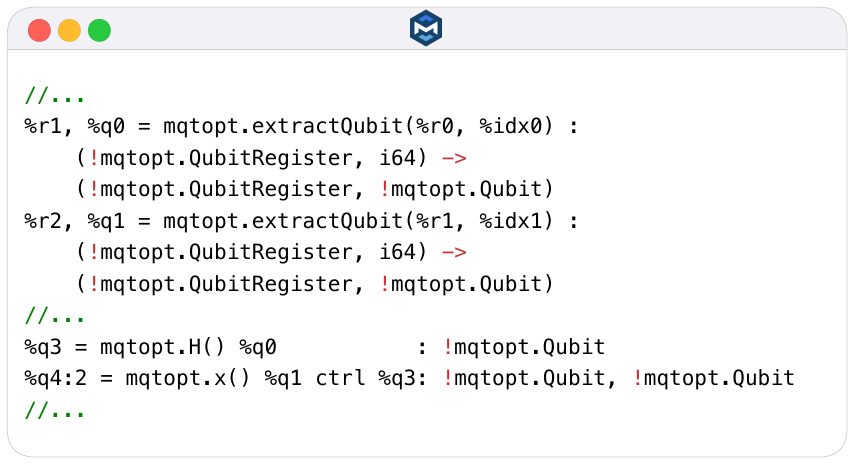}
      \caption{\texttt{MQTOpt} MLIR dialect}
      \Description{Subfigure showing the MQTOpt dialect with dedicated operations and types, illustrating a QubitRegister type and an example gate operation with attributes.}
      \label{fig:mqtopt}
  \end{subfigure}
  \caption{Two MLIR dialects representing (parts of) the same quantum program.}
  \Description{Two-panel figure comparing the Catalyst Quantum dialect and the MQTOpt dialect, highlighting differences in gate representation, types, and operand/result conventions.}
  \label{fig:quantum-mqtopt-sidebysidea}
\end{figure*}

\subsection{Implementation}

Integrating PennyLane and MQT into MLIR requires several foundational components.  
To facilitate a clear understanding of these building blocks, we present a dedicated MLIR \emph{dialect}, associated \emph{transformations}, and the embedding \emph{plugin} infrastructure.  
We encourage readers to explore the accompanying open-source implementation available in the \href{https://github.com/munich-quantum-toolkit}{MQT GitHub repository}:

\begin{itemize}  %
  \item[\faGithub] \small{\href{https://github.com/munich-quantum-toolkit/core-plugins-catalyst}{https://github.com/munich-quantum-toolkit/core-plugins-catalyst}}
  \item[\faBook] \small{\href{https://mqt.readthedocs.io/projects/core-plugins-catalyst}{https://mqt.readthedocs.io/projects/core-plugins-catalyst}}
\end{itemize}

\textbf{Dialect:} 
The \texttt{Quantum} MLIR dialect provided by {Xanadu}'s Catalyst enables direct access to the structure and semantics of quantum programs.
For example, parts of the MLIR representation of the GHZ circuit implemented in \autoref{fig:plugin-mapping} are shown in \autoref{fig:catalyst-quantum} using Catalyst's SSA format.

However, as outlined in \autoref{sec:background}, many quantum software tools do not support MLIR out of the box.
Since the MQT similarly lacks native MLIR support, a translation between the MQT Core IR and MLIR is necessary.

To ease this transition and minimize the disruption caused by such a change, a dedicated quantum dialect (called \texttt{MQTOpt}) has been created that very closely matches the semantics of the existing MQT Core IR. 
In addition to SSA semantics, it enforces a single-use constraint, where each value is consumed only once.
Such a linear typing property naturally enforces the no-cloning theorem~\cite{Wootters_1982}, as each qubit is guaranteed to be defined and used exactly once.
This further simplifies the tracking of register state changes during qubit extraction or insertion, thereby easing subsequent translations to and from program representations that explicitly model physical qubit assignments on hardware.
For instance, extracting a qubit from a quantum register yields both a qubit and an updated register, and is represented in the custom \texttt{MQTOpt} dialect as illustrated in \autoref{fig:mqtopt}.

Moreover, \autoref{fig:mqtopt} highlights dialect-specific operand and result \emph{types} (e.g., the \texttt{QubitRegister}) and \emph{operations} (e.g., the \texttt{x}-gate), along with an associated modifier (here, \texttt{ctrl}).
Another key difference from the \texttt{Quantum} dialect (cf. \autoref{fig:catalyst-quantum}) is that, while Catalyst represents gates as generic operations distinguished by a \texttt{gate\_name} attribute, the \texttt{MQTOpt} dialect defines each gate as a dedicated, dialect-specific operation.
Note that the code shown is only a textual representation of the program (automatically produced by MLIR) and does not necessarily align with the names of classes, types, and attributes in the C++ implementation.

The implementation of a dialect can be further streamlined by defining \emph{traits}, which encapsulate common properties shared across different types or operations.
In the context of the \texttt{MQTOpt} dialect, structural invariants---such as the expected number of parameters for a given operation (e.g., a rotation gate)---are enforced by implementing a \texttt{verifyTrait()} function within the corresponding trait definition.

\textbf{Transformation:} 
Next, to enable seamless translation between the two dialects, MLIR transformation patterns offer an effective and convenient mechanism for defining translations at different levels of abstraction.
In general, three types of MLIR transformations can be distinguished: Importing/exporting to/from non-MLIR representations, conversions between different MLIR dialects, and transformations within a single dialect.
Specifically, conversions from higher-level dialects or IRs to lower-level dialects are referred to as lowerings.

In the context of the integration, a (partial) conversion was implemented by providing a \texttt{TypeConverter} along with several \texttt{ConversionPattern}s to translate between the source and target quantum dialects (leaving non-quantum logic untouched).  
In contrast to the dialect definitions, which largely consist of boilerplate code, such patterns typically need to be implemented manually, as they often involve non-trivial rewrite logic---for example, enforcing the single-use constraint in the \texttt{MQTOpt} dialect.

At this point, it is important to note that having implemented a dialect and its associated conversions has already established all the necessary components to support compilation of quantum programs within the MLIR framework.  
With these components in place, future development can---but does not have to---build upon MLIR's pass infrastructure to enable advanced compiler analyses and transformations.

\begin{example}
To implement a qubit mapping routine, one may define analysis passes to extract qubit interaction patterns, followed by transformation passes that minimize the number of required \texttt{SWAP} operations.  
MLIR's pass infrastructure supports runtime configurability through \texttt{PassOptions}, which enable, for example, the specification of a coupling map via key-value pairs, as illustrated in \autoref{fig:plugin-mapping}.
\end{example}

\begin{figure*}[t]
    \vspace{10pt}
    \centering
    \begin{subfigure}{0.49\linewidth}
        \centering
        \includegraphics[width=\linewidth]{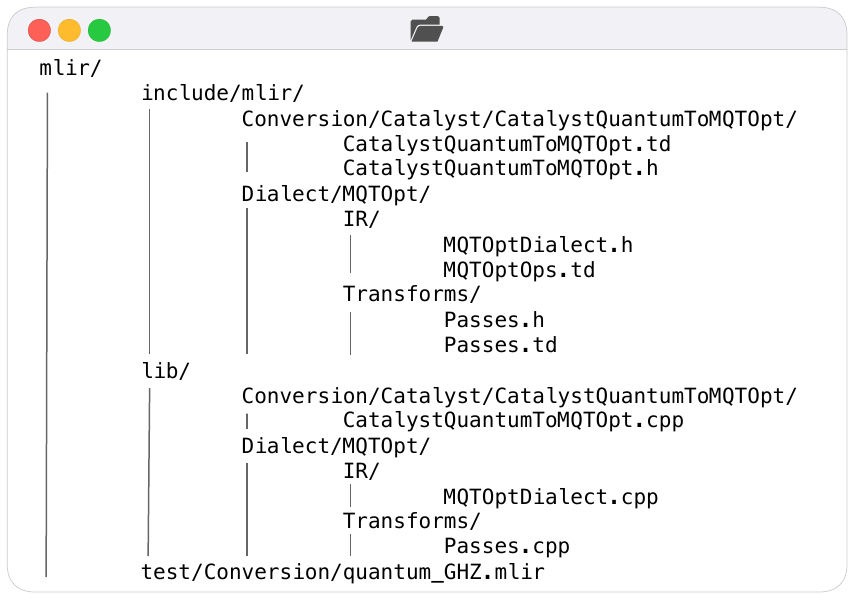}
    \caption{\vspace{7pt}MQT's MLIR project structure.}
    \Description{Illustration of the repository layout used for the MQT MLIR integration, showing directories for ops, passes, TableGen files, tests, and build scripts.}
    \label{fig:project-structure}
    \end{subfigure}
    \hfill
    \begin{subfigure}{0.49\linewidth}
        \centering
        \includegraphics[width=\linewidth]{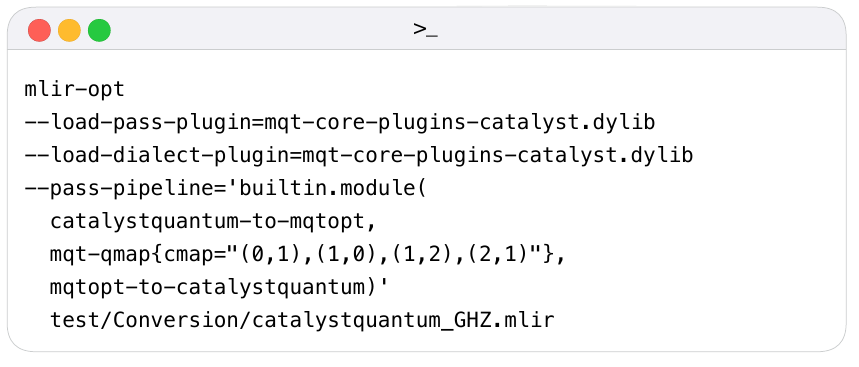}
    \caption{\vspace{7pt}Executing a mapping round trip with an MLIR plugin.}
    \Description{Sequence diagram showing a mapping round trip where an MLIR plugin imports a program, runs mapping passes, exports results, and validates the transformed IR. \vspace{10pt}}
    \label{fig:roundtrip}
    \end{subfigure}
    \caption{Overview of a typical file structure of an MLIR-based project and its \texttt{mlir-opt} usage.}
  \Description{Composite figure summarizing the Catalyst and MQTOpt dialects and related tooling, illustrating how the components interact during import, transformation, and verification.}
    \label{fig:roundtripall}
  \vspace{5pt}
\end{figure*}

\textbf{Plugin:} 
Ultimately, our goal is to make the MLIR passes defined above available without requiring users to compile the entire library---or, more broadly, the entire MLIR project.
Fortunately, MLIR's pass manager infrastructure supports the registration of passes and the dynamic loading of plugin-based dialects and passes.
To enable this dynamic approach, one must compile an MLIR \texttt{DialectPlugin} and \texttt{PassPlugin} into a shared library.
The compiled files can then, for example, be distributed as pre-compiled binary wheels similar to traditional compiled extensions.

Similarly, tool developers do not have to compile the entire Catalyst project in order to implement a dedicated plugin.
In the context of its MQT integration, the implementation only requires the \texttt{Quantum} dialect header files instead of having to build the entire library.

Finally, by leveraging the shared library plugin, the pass pipeline---specifying all relevant passes required for the full round trip---can be executed.
The definition of a custom dialect, the transformations to and from this dialect, and the accompanying plugin infrastructure provide all the essential building blocks for integrating quantum software tools within the MLIR ecosystem.

\begin{example}\label{ex:roundtrip}
Consider, once more, the task of mapping a quantum program defined in PennyLane using QMAP---i.e., performing a full round trip between Catalyst and the MQT.  
An example invocation of the complete round-trip pipeline is illustrated in \autoref{fig:roundtrip}.  
Starting from the GHZ circuit defined in the \texttt{Quantum} dialect (shown in \autoref{fig:catalyst-quantum} and provided in \texttt{quantumGHZ.mlir}), the conversion to \texttt{MQTOpt} and back is handled by the \texttt{catalystquantum-to-mqtopt} and \texttt{mqtopt-to-catalystquantum} passes, respectively.  
Mapping via QMAP, along with export to and re-import from the MQT Core IR, is performed using the \texttt{mqt-qmap} pass.  
Note that the coupling map is specified using the \texttt{cmap} pass option.  
The entire round trip can be executed using the \texttt{mlir-opt} command-line utility, which supports loading dialects and pass plugins via shared library files.
\end{example}

\section{Best Practices}

After gaining familiarity with the fundamental building blocks of a representative MLIR integration, the initially high entry barrier has already been lowered.  
To further ease the adoption process, we present a set of best practices that have proven essential for the successful realization of the integration and can greatly reduce the development effort of similar projects. 
Quantum software developers seeking to adopt MLIR in their own tooling are encouraged to take these into account. 

\emph{First, do not reinvent the wheel.}
MLIR is a rapidly growing ecosystem, supported by contributions from both industry and academia, and includes numerous well-maintained open-source projects.  
To avoid duplicating existing solutions, it is advisable to engage with the broader MLIR community---particularly through platforms like Discord or GitHub---and to study existing implementations.
Numerous MLIR dialects, such as the arithmetic dialect \href{https://mlir.llvm.org/docs/Dialects/ArithOps/}{\texttt{arith}} and the structured control flow dialect \href{https://mlir.llvm.org/docs/Dialects/SCFDialect/}{\texttt{scf}}, provide sophisticated, reusable components that can be directly adopted or extended in new projects.  
In the spirit of encouraging reuse, we have made the implementation of the MLIR dialect and plugin described in this paper available as open source in the \href{https://github.com/munich-quantum-toolkit}{MQT GitHub repository}.

The project structure is illustrated in \autoref{fig:project-structure}, and it follows the directory layout commonly found in MLIR-based projects.
Gaining familiarity with existing projects helps prevent redundant development efforts and provides valuable insights into dialect design and implementation.

\emph{Secondly, prioritize modular and lightweight design.}
Without careful design choices, quantum software engineers---who are often not deeply familiar with the intricacies of LLVM or MLIR---can easily be overwhelmed by their size and complexity (the very reason for writing this article).  
Modularity---a foundational principle in modern compiler development---helps mitigate their overhead by promoting maintainability, testability, and extensibility in increasingly complex software ecosystems.
In practice, this means isolating dependencies and avoiding unnecessary compilation of massive codebases like the entire LLVM project.  
For example, as full-stack frameworks such as PennyLane continue to grow, compiling or linking against large monolithic toolchains becomes increasingly impractical.  
Fortunately, MLIR's pass and plugin infrastructure allows developers to cleanly encapsulate custom functionality without requiring deep knowledge or changes to the core system.
In the integration effort presented above, we therefore provided the \texttt{MQTOpt} dialect and its associated passes as a precompiled MLIR plugin.
Embracing modular, lightweight, and plugin-based architectures improves maintainability and lowers the barrier to entry for new contributors.
While some situations (like, e.g., debugging and testing) may still require building MLIR (or other large dependencies) from source, a well-isolated design ensures this remains an exception rather than the norm.

\emph{Finally, use dedicated tooling.}
LLVM and MLIR provide a wide range of utilities that can greatly simplify the development of quantum software tools.  
Among these, two tools are particularly helpful and are used throughout the open-source repository accompanying this paper. \\ 
Rather than providing a full tutorial, our goal here is to highlight their purpose and role in the development process.

\begin{figure*}
  \centering
  \includegraphics[width=0.98\linewidth]{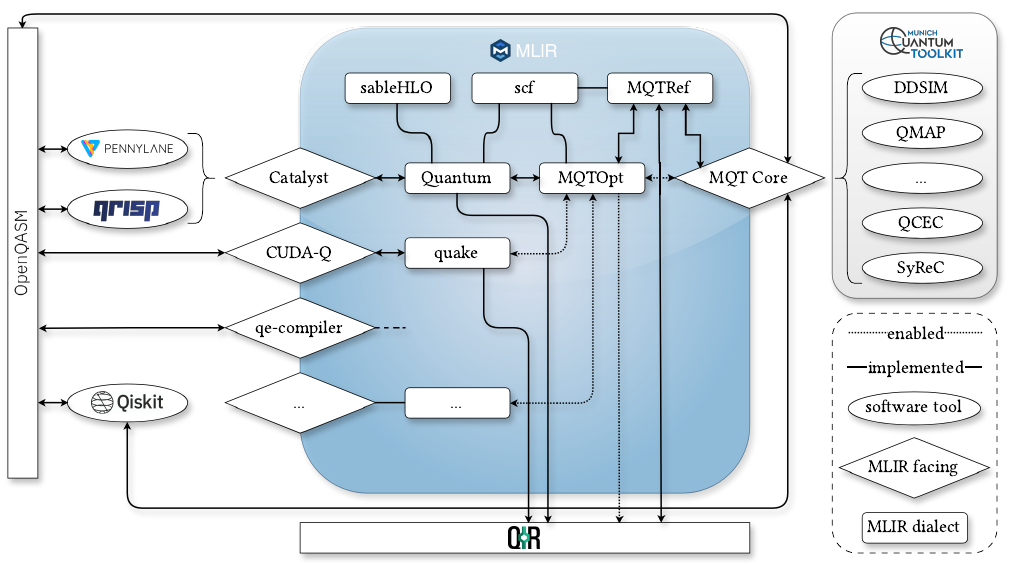}
  \caption{Map of quantum computing software tools within and outside of the MLIR ecosystem.}
  \Description{Graphic map placing various quantum software tools and frameworks inside or outside the MLIR ecosystem, grouping logos and arrows to show integration paths and missing links.}
  \label{fig:overview}
\end{figure*}

First, the implementation of key MLIR components can be significantly streamlined using the so-called \href{https://llvm.org/docs/TableGen/}{\texttt{TableGen}} tool and its associated definition files.  
These files enable the automatic generation of much of the boilerplate code required by MLIR, including header declarations and getter/setter methods---supporting implementations that range from highly general to fully customized (i.e., manually defined).
As illustrated in \autoref{fig:project-structure}, the operations in the \texttt{MQTOpt} dialect are defined in the corresponding \texttt{MQTOptOps.td} file (following naming conventions in the MLIR ecosystem).  
While this definition file contains fewer than 700 lines of code, the corresponding \texttt{MQTOptOps.cpp.inc} file generated by TableGen comprises nearly 20,000 lines of complicated template code---demonstrating the considerable implementation effort that TableGen abstracts away.

Second, \href{https://mlir.llvm.org/docs/Tutorials/MlirOpt/}{\texttt{mlir-opt}} (already briefly touched on in \autoref{ex:roundtrip}) provides a centralized driver that connects and orchestrates the execution of MLIR passes.  
Its command-line entry point is particularly useful during development and debugging, allowing fine-grained inspection of the IR at different compilation stages.  
Combined with the LLVM tools \href{https://llvm.org/docs/CommandGuide/lit.html}{\texttt{lit}} and \href{https://llvm.org/docs/CommandGuide/FileCheck.html}{\texttt{FileCheck}}, it can also facilitate the systematic testing of MLIR passes and transformations.
In the context of the PennyLane–MQT MLIR integration, dedicated test files (e.g., \texttt{quantumGHZ.mlir}) are annotated with \texttt{lit} commands and \texttt{FileCheck} directives that run \texttt{mlir-opt} and verify the correctness of its output after specific transformations have been applied.
In summary, dedicated tools like \texttt{TableGen} and \texttt{mlir-opt} not only reduce manual effort and potential implementation errors but also promote consistency across the codebase and enable the application of automated testing workflows.

Adhering to these best practices promotes an effective and sustainable MLIR integration, whose immediate and long-term advantages are discussed next.

\section{Discussion}\label{sec:discussion}

The sections above have outlined how the integration of quantum software tools within the MLIR framework can be efficiently achieved, thereby significantly lowering the entry barrier for quantum software developers.
Next, key insights gained throughout the integration, as well as directions for future work, are discussed.  
While the following highlights are drawn from the MQT integration, the insights are broadly applicable to any quantum software tool aiming to adopt MLIR.

\subsection{Insights}

With the key components and best practices for a successful MLIR integration established, we now take a step back to evaluate the \emph{immediate} implications of this effort.
To this end, consider \autoref{fig:overview}, which illustrates the newly achieved interaction between Catalyst and MQT Core, facilitated entirely within the MLIR framework.  

The integrated tool is now part of a largely community-developed ecosystem, joining a growing family of MLIR adopters.  
As a result, it benefits from ongoing support and advancements across the MLIR ecosystem.  
In particular, performance improvements and updates by experts from the MLIR community directly enhance the performance of tools that utilize the MLIR infrastructure.

Another advantage is the improved usability for users on both sides---those working with PennyLane as well as those using the MQT.
Quantum algorithm experts, for example, can now focus on implementing their algorithms using the straightforward Catalyst decorators for mapping with QMAP, as illustrated in \autoref{fig:plugin-mapping}, rather than navigating the cumbersome workflow depicted in \autoref{fig:current_workflow}. 

Leveraging the new integration, the interaction between Catalyst and MQT Core enables interoperability between third-party quantum computing tools that were previously only connected through text-based exchange formats such as OpenQASM.
On one side, tools that build upon the Catalyst compiler, such as Qrisp~\cite{seidel2024qrispframeworkcompilablehighlevel}, can now directly interface with the MQT compilation tools.
While this has not yet been tested at the time of writing, it is expected to require minimal effort.  
On the other end, any tool built on top of the MQT Core library---such as DDSIM~\cite{Burgholzer2022} or QCEC~\cite{Peham_2023}---can now consume MLIR input. 
This connection directly facilitates interoperability across these software tools on the compiled level, significantly expanding the landscape of composable quantum compilation workflows.

\subsection{Outlook}

Taking an even broader perspective on \autoref{fig:overview}, additional \emph{long-term} advantages of an MLIR-based integration become evident.
Specifically, the integration enables easy extensibility to and seamless interoperability with a broad range of quantum software tools, and paves the way for a more general integration into established software stacks.
For instance, integration with other MLIR-based frameworks---such as {NVIDIA}'s CUDA-Q---can now be achieved with minimal additional effort.  
In such a scenario, little more than conversion passes between the \texttt{MQTOpt} dialect and CUDA-Q's \texttt{quake} dialect would need to be implemented---an extension made low-effort by the already existing and modular code base.

Furthermore, the integration of classical dialects, such as \texttt{scf}, can now be conducted with ease.  
Catalyst's tracing capabilities, for example, already benefit significantly from structured control flow constructs like loops, as provided by \texttt{scf}.
Extending the \texttt{MQTOpt} dialect to support structured control flow will enable more compact program representations and improve the applicability of compilation passes.  
For instance, quantum circuits composed of repeated patterns can be expressed with constant program size, an advantage that becomes especially relevant in nontrivial instances of algorithms such as Grover's.  
Moreover, in these cases, it may suffice to map the fixed qubit interactions within a single loop iteration, rather than unrolling the entire circuit and redundantly mapping identical patches---common, e.g., in quantum machine learning circuits---multiple times.

More broadly, this observation holds for many quantum software tools considering future MLIR integration.  
Tasks such as circuit analysis and subsequent transformations---often computationally intensive and typically implemented in Python---stand to benefit substantially from the efficient, structured program representation that MLIR provides.  
By shifting such functionality to the MLIR layer, significant performance gains and improved scalability can be achieved.

Another observation evident from \autoref{fig:overview} is that Qiskit has thus far remained relatively isolated from the MLIR ecosystem. 
The associated \texttt{qe-compiler}~\cite{healy2024designarchitectureibmquantum}, which converts OpenQASM into several dedicated MLIR dialects, has not seen widespread adoption and remains largely unmaintained.
At the same time, MQT Core already supports direct integration with Qiskit at the Python level. 
With Qiskit's ongoing transition to a Rust-based backend, directly connecting the MLIR and Qiskit ecosystems presents an interesting challenge. 
The recent development of a \href{https://docs.quantum.ibm.com/api/qiskit-c}{C API for Qiskit}'s compiled components may open up a viable path forward, making MQT Core a promising candidate for enabling integration between the two ecosystems.

Finally, the demonstrated MLIR plugin paves the road to a unified interaction with QIR as an exchange format.  
Because QIR is built on top of LLVM IR, it is already being utilized for deploying quantum accelerators in HPC software stacks~\cite{Shehata_2026}.  
For the same reason, QIR is familiar to MLIR and can easily be integrated with an MLIR-based compilation flow.
As a result, MLIR-integrated tools, such as MQT, can be applied directly to QIR files.  
Conversely, any tool that already supports QIR can be readily incorporated into the flow, contributing to a modular and extensible quantum software stack.

In summary, the immediate and long-term benefits demonstrate that lowering the entry barrier for quantum software engineers seeking to adopt the MLIR framework is a worthwhile endeavor.
Their integration efforts will significantly enhance interoperability and lay the foundation for a more interconnected quantum computing ecosystem, ultimately paving the way for its adoption into classical and HPC software stacks.

\newpage
\section{Conclusions}\label{sec:conclusion}

This work demonstrated how to leverage MLIR's dialect and pass plugin system to develop a modular and lightweight compilation backbone---building upon decades of established classical compiler expertise.  
To illustrate this, we presented a practical example of integrating two major quantum computing frameworks, replacing previously cumbersome and ad hoc parsing mechanisms with a streamlined MLIR-based solution. 
Recognizing the steep learning curve posed by MLIR for quantum software developers, we provided best practices and insights derived from this integration effort.
The resulting implementation is provided as open-source software, intended as both a foundation and inspiration for similar integration efforts within the broader quantum computing community.
Ultimately, this work provides essential guidance for quantum software engineers who are unfamiliar with MLIR and wish to integrate their tools with(in) the framework and thereby improve interoperability with other software tools, as well as with classical and HPC infrastructures.
In doing so, it significantly lowers the entry barrier traditionally associated with adopting MLIR for quantum compilation and paves the way for its integration into established HPC software stacks.

\begin{acks} %
Authors affiliated with the Technical University of Munich acknowledge funding from the European Research Council (ERC) under the European Union’s Horizon 2020 research and innovation program (DA QC, grant agreement No. 101001318 and MILLENION, grant agreement No. 101114305), the Munich Quantum Valley, which is supported by the Bavarian state government with funds from the Hightech Agenda Bayern Plus.
This work has been supported by the BMK, BMDW, the State of Upper Austria in the frame of the COMET program, the BMFTR within the FullStaQD inititive and on the basis of a decision by the German Bundestag through the project SYNQ, as well as the Deutsche Forschungsgemeinschaft (DFG, German Research Foundation – 563436708).

Xanadu authors thank the Catalyst and PennyLane teams at Xanadu for useful discussions and reviews of the MLIR plugin features of Catalyst that made this implementation possible, including but not limited to David Ittah, Mehrdad Malek, Paul Haochen Wang, Joey Carter, and Raul Torres. All authors thank the MLIR and LLVM contributors for all their hard work.
\end{acks}

\newpage

\clubpenalty=10000
\widowpenalty=10000
\interlinepenalty=10000

\bibliographystyle{ACM-Reference-Format}
\bibliography{references}

\end{document}